\documentclass[twocolumn,pre,showpacs,floatfix]{revtex4}
\usepackage{epsfig}
\usepackage{amsfonts}
\usepackage{amssymb}
\usepackage{amsmath}
\usepackage{pgf,pgfarrows,pgfnodes,pgfautomata,pgfheaps,pgfshade}
\usepackage{graphics,graphicx}

\begin{document}
\title{Embedded solitons in the double sine-Gordon lattice 
with the next neighbor interactions}
\author{Yaroslav Zolotaryuk}
\email{yzolo@bitp.kiev.ua}
\affiliation{Bogolyubov Institute for Theoretical Physics,
National Academy of Sciences of Ukraine,
vul. Metrologichna 14B, 
 03143 Kyiv, Ukraine}
\author{Ivan O. Starodub}
\email{starodub@bitp.kiev.ua}
\affiliation{Bogolyubov Institute for Theoretical Physics,
National Academy of Sciences of Ukraine,
vul. Metrologichna 14B, 
 03143 Kyiv, Ukraine} 

\date{\today}

\begin{abstract}
Topological solitons can propagate without radiation in discrete
media. These solutions are known as embedded solitons (ES). They
come as isolated solutions and exist despite their resonance with
the linear spectrum of the respective lattices. 
In this paper the properties of embedded solitons in the discrete 
double sine-Gordon equation with the next-neighbor and second-neighbor 
interactions are investigated. Depending on the sign of these
interactions they can be either destructive or favorable for the ES 
creation.
The ES existence area depends on the width of the linear spectrum: 
narrowing of the spectrum widens the ES existence range
and vice versa.
The application to the Josephson junction arrays is discussed.

\end{abstract}
\maketitle

%####################################################################
%####################################################################
\section{Introduction} \label{intro}
%####################################################################
%####################################################################

The double sine-Gordon (DbSG) equation \cite{cgm83prb,cps86pd} is used in 
many physical systems, including ultrashort optical pulses that 
propagate in degenerate media \cite{dbd75jpa}, spin waves 
in superfluid $^3$He \cite{mk76prb},
nonlinear waves in the piezoelectric XY model \cite{r81jpc}. 
It also serves as an approximation of the non-sinusoidal 
generalizaitons of the Frenkel-Kontorova model \cite{pr82prb,bkz90prb}.
In particular, we would like to highlight the applications of
the DbSG equation to the systems based on the Josephson effect.
It is used to describe the dynamics of long Josephson junctions (JJs)
of the superconducotor-ferromagnet-superconducotor (SFS) and/or
superconducotor-insulator-ferromagnet-superconducotor (SIFS) type
\cite{gkkb07prb,abszs10jpcs}. In these junctions the current-phase relation
differs significantly from the single-harmonic dependence and the
second harmonic is taken into account\cite{gki04rmp,a15ltp}.
Also the non-local generalization of the DbSG has been used to 
describe the long JJ where the superconducting layers are thin \cite{amm14pd}.
The discrete version of the DbSG equation has been introduced in 
Ref. \cite{nknfh10pc} for the asymmetric array of JJ SQUIDs
(superconducting quantum interference devices).
An important feature of the spatially distributed JJ systems (both
continuous and discrete) is existence of the topological solitons. They
carry a magnetic flux quantum and are known as fluxons or Josephson
vortices \cite{barone82,u98pd}.

One important property of the discrete double sine-Gordon (DDbSG) 
equation is that it possesses \cite{zs15pre}
moving {\it embedded} solitons (ESs). Embedded solitons \cite{cmyk01pd} 
are solitons that
exist in non-integrable systems and are in resonance with the linear
waves of these systems. In particular, for the discrete media
that are modelled by the equations of the nonlinear Klein-Gordon (NKG) type
[discrete sine-Gordon (DSG), DDbSG, $\phi^4$] this resonance is 
the resonance between
the soliton velocity and the phase velocity of the linear waves.
As was first pointed out in Ref. \cite{pk84pd} this happens because
for the DNKG type equations the linear spectrum always
has a gap, thus for any soliton velocity $v$ there is at least one
non-zero root of the equation $vq=\omega_L(q)$, where $\omega_L(q)$
is the linear wave spectrum and $q$ is the wavenumber. 
As a result any propagating soliton must
be accompanied with the linear wave with the same phase velocity $v$. 
These solutions with the oscillating tails 
are known in the literature as {\it nanopterons} \cite{b90no}.
Nevertheless, such a resonance can
be avoided if there is only one root of the above-mentioned
equation \cite{acr03pd}. Note that acoustic lattices
with the gapless spectrum do not have this problem and have
a continuous velocity spectrum for the moving solitons.
\cite{fw94cmp}. 
In some cases the embedded soliton can be found
explicitly \cite{s79prb} and, moreover, systems that support
embedded solitons can be generated in a systematic way \cite{fzk99pre}.
A number of analytical \cite{bop05pre,opb06n,amp14prl} and numerical
\cite{zes97pd,sze00pd,kzcz02pre,dkksh08pre,akjs-mg-r13ujp} has demonstrated 
that discrete ESs
are not a isolated effect but a generic phenomenon that occurs in 
various lattice models with the different physical background. 
The existence of ESs has been shown experimentally in the JJ
arrays (JJAs) \cite{psau06prl}.
In this case the ESs were the bound states of two or more DSG kinks
(fluxons) that propagate with velocities that are significantly different
from the individual kink velocity. Their existence manifests itself
as a distinct branch of the current-voltage curve of the array.
Continuous ESs  have been demonstrated to exist in the DbSG with the
fourth order dispersion \cite{bkm01wm} and with the non-local 
dispersion \cite{amm14pd}.

The more correct description of the various nonlinear phenomena in lattices
requires consideration of not just the nearest-neighbor interactions but
also the next-neighbor and/or the further distant neighbor interactions
\cite{bkz90prb,gfnm95prl,skdhpnt09ol,cake18pre}. In the case of JJ arrays 
this means that
not only the coupling due to the self-inductance of each JJ cell should be
accounted for, but also the mutual inductances between the cells  \cite{pzwo93prb}
should be taken into consideration.
In this paper our aim it to study how the presence of the next-to-nearest
interactions influences the properties of ESs.

The paper is organized as follows. The model of the Josephson junction array 
and the equations of motion are given in the next section. The 
linear spectrum is defined in Sec.\ref{disp}. In Sec. \ref{ham} we 
discuss the properties of the
JJA in the hamiltionian (dissipationless) limit. Next section is devoted to the
current-voltage characteristics. Discussion and conclusions are
given in the last section.

%####################################################################
%####################################################################
\section{The model and equations of motion}\label{model}
%####################################################################
%####################################################################

Here we study the resistively and capacitatively shunted 
array model (RCSJ) of the
small SFS or SIFS junctions where the intercell inductance is taken
into account. According to \cite{gki04rmp,a15ltp} for such junctions
one should consider
not only the first harmonic of the current-phase relation, but also
the second one: $I_s(\phi)=I_c \sin \phi+ I_c^{(2)} \sin 2\phi$.
In the RCSJ model the equations of motion of the JJA are derived
from the combined Josephson relations, the Kirchhoff law and
the flux quantization rules 
\cite{wzso96pd,u98pd}. The main dynamical variable is the Josephson phase
$\phi_n$
of the $n$th junction which is the difference of the phases
of the wavefunctions of the superconducting electrodes that form
tha junction. 
Below we write down the equations of motion:
%--------------------------------------------------------------------
\begin{eqnarray} \nonumber
&&C\frac{\hbar}{2e} \ddot{\phi_n}+\frac{\hbar}{2eR} \dot{\phi_n}+
I_c (\sin \phi_n +\eta \sin 2\phi_n)= \\
\label{1}
&& =I_B+I_{n-1}-I_n,~~~\eta=\frac{I_c^{(2)}}{I_c},\\
\label{2}
&&\phi_{n+1}-\phi_n=-\frac{2\pi}{\Phi_0} \left[ L_{0} I_n+
L_{1} (I_{n-1}+I_{n+1}) \right ]~,
\end{eqnarray}
%--------------------------------------------------------------------
where $C$ is the cell capacitance, $R$ is the junction resistance, $I_c$
is its the critical current, $\Phi_0=\pi\hbar/e$ is the magnetic flux
quantum. The current $I_n$ is the 
mesh current flowing in the $n$th cell of the array. In our notations
the $n$th cell is placed between the $n$th and $(n+1)$th junctions.
The dimensionless parameter $\eta$ measures the value of the
second harmonic in the current-phase relation in the units of the
main harmonic. According to the previous research \cite{gkkb07prb} it
can change in the very broad range of values, both negative and positive.
In this paper we will stick to the positive $\eta$.

The more accurate physical approach is to take into 
account not only the 
self-inductance of the junction cell, but also the mutual
inductances between all the cells \cite{pzwo93prb,dj96prb}. 
As a result, the flux through
the $n$th cell should depend on all currents as $\Phi_n=\sum_n L_{mn}
I_m$, where $L_{mn}$ are the elements of the inductance matrix. The
diagonal element of this matrix is the self-inductance coefficient,  
and the off-diagonal elements are the mutual inductance coefficients.
The properties of the inductance matrix has been studied in detail 
in the number of papers \cite{dj96prb,mc96prb,fpr99epjb}. Because 
the mutual inductance coefficient between the $n$th and $m$th cells
decays as $L_{mn} \propto |m-n|^{-3}$ we limit ourselves to the 
mutual inductances between the neighboring cells. Thus, we denote 
the self-inductance as $L_{nn}=L_0>0$, and the mutual inductance between
the neighboring cells as $L_{n,n\pm 1}=L_1$. Usually, the 
mutual inductance in JJAs is negative \cite{dj96prb}, however 
may be positive under some special current properties \cite{mc96prb}.
If only the cell self-inductance is taken
into account, the Eqs. (\ref{1})-(\ref{2}) are easily reduced to the
discrete double sine-Gordon (DDbSG) equation with the nearest neighbor couplings
\cite{wzso96pd}. 

We consider here the circular JJA's, thus, the periodic boundary conditions
should apply. The curvature effects are neglected if the number of junctions
in the array $N$ is large.

The second of Eqs. (\ref{1})-(\ref{2}) can be rewritten in the matrix 
form
%--------------------------------------------------------------------
\begin{eqnarray}\label{3}
&&\hat {S} \vec {\phi} = - \frac{2\pi L_{0}}{\Phi_0} \hat{\Lambda} \vec I ,\\
&&\vec {\phi} = (\phi_1,\phi_2,\ldots,\phi_N)^T,~
\vec {I}= (I_1,I_2,\ldots,I_N)^T~,
\end{eqnarray}
%--------------------------------------------------------------------
where two $N\times N$ {\it circulant} \cite{g06} matrices appear. 
The matrix $\hat S$ is bidiagonal 
%--------------------------------------------------------------------
\begin{equation}
\hat {S}=\left (\begin{array}{cccccc}
        -1 &  1   & 0      &  0     & \cdots &0      \\
         0 & -1   & 1      &  0     & \cdots &0      \\
    \cdots &\cdots& \cdots & \cdots & \cdots &\cdots \\  
        0  &  0   &   0    & \cdots & -1     &1 \\  
         1 &  0   & 0      & \cdots &  0     &-1 
         \end{array} \right )~,
\end{equation}
%--------------------------------------------------------------------
and the dimensionless 
inductance matrix $\hat \Lambda$ is  symmetric 
%--------------------------------------------------------------------
\begin{equation}\label{6}
\hat {\Lambda}=\left (\begin{array}{cccccc}
       1   & \nu   & 0      & \cdots   & 0      &\nu     \\
       \nu & 1     & \nu    & \ddots   & 0      &0       \\
       0   & \nu   & 1      & \ddots   & 0      &0       \\
    \vdots &\ddots & \ddots & \ddots   & \ddots &\vdots  \\  
         0 &  0    & 0      &   \nu    &  1     & \nu    \\   
       \nu &  0    & 0      & \cdots   & \nu    & 1 
         \end{array} \right )~.
\end{equation}
%--------------------------------------------------------------------
These matrices are circulant due to the periodicity of the boundary
conditions. For the linear array they would be standard Toeplitz matrices.
The parameter in matrix $\hat \Lambda$ is the ratio 
$\nu=L_1/L_{0},~|\nu|< 1$.
It is desirable to express the mesh currents $I_n$ as a function
of the phases $\phi_n$ and substitute them into 
Eqs. (\ref{1})-(\ref{2}). The 
circulant matrix (\ref{6}) can be inverted using the known techniques \cite{g06}.
As a result, we obtain the elements of the inverted matrix
%--------------------------------------------------------------------
\begin{equation}
\Lambda^{-1}_{mn}=\frac{1}{N}\sum_{k=1}^N 
\frac{e^{\i \frac{2\pi}{N}(k-1)(n-m) }}{1+2\nu\cos 
\left(2\pi \frac{k-1}{N}\right)}~, 
\end{equation}
%--------------------------------------------------------------------
where we keep in mind that $\nu$ is a small parameter. Next, we
expand the elements of $\hat {\Lambda}^{-1}$ into the Taylor series
with respect to the powers of $\nu$. If we ignore the terms
smaller than ${\cal O}(\nu^k)$, the matrix $\hat{\Lambda}^{-1}$
would become circulant with $2k+1$ non-zero diagonals. For example,
if we keep only the linear terms, we would have a 
tridiagonal circulant matrix,
if the ${\cal O} (\nu^2)$ terms are included the inverse matrix would 
become pentadiagonal:
%--------------------------------------------------------------------
\begin{equation}\label{8}
 \Lambda^{-1}_{mn}=\left \{\begin{array}{cc}
              1+2\nu^2,& m=n;\\
              -\nu,    & m=n\pm 1; \\
                       & m=1, n=N; \\ 
                       & m=N, n=1; \\
              \nu^2,   & m=n\pm 2; \\
                       & m=1, n=N-1; \\
                       &  m=2, n=N; \\
                       & m=N, n=N-1; \\
                       & m=N-1, n=2; \\
              0,       & \mbox{else}
             \end{array} \right . \;\;\; +{\cal O}(\nu^3).
\end{equation}
%--------------------------------------------------------------------
We shall limit ourselves with the ${\cal O} (\nu^2)$ terms.
Now the inverse matrix $\hat {\Lambda}^{-1}$ should be substituted 
into Eq. (\ref{3}). After that it becomes possible to express the mesh 
currents through the Josephson phases explicitly:
%--------------------------------------------------------------------
\begin{eqnarray}\label{9}
&&I_n=-\frac{\Phi_0}{2\pi L_{0}}\left [\nu^2 (\phi_{n+3}-\phi_{n-2})+
\right . \\
&& \nonumber \left . 
+\nu(1+\nu)(-\phi_{n+2}+\phi_{n-1})+ \right .\\
&& \nonumber \left .+(1+\nu+2\nu^2)(\phi_{n+1}-\phi_n) 
+ {\cal O}(\nu^3)\right ],\; n=1,2,\ldots N.
\end{eqnarray}
%--------------------------------------------------------------------
This expansion is substituted into Eq. (\ref{1}) and the terms
of the order ${\cal O}({\nu^3})$ and weaker are neglected.
After introducing the dimensionless variables
%--------------------------------------------------------------------
\begin{eqnarray}
&&t\to t\omega_J,\;\omega_J=\sqrt{\frac{2eI_c}{C\hbar}},\;\\
&& \nonumber
\alpha=\frac{\hbar\omega_J}{2eI_cR},\;
\gamma=\frac{I_B}{I_c},\; \kappa  =\frac{\Phi_0}{2\pi L_0 I_c}~,
\end{eqnarray}
%--------------------------------------------------------------------
one arrives to the DDbSG equation with the
next-to-nearest and second-to nearest neighbor interactions:
%--------------------------------------------------------------------
\begin{eqnarray} \label{11}
&&\ddot{\phi}_n - \kappa\, \left [\sum_{j=1}^3 D_j(\nu)\hat{\Delta}_j 
\right] \phi_n+ \alpha  \dot{\phi}_n+
\\ 
\nonumber
&&+ \sin {\phi_n}+\eta\sin {2\phi_n}=\gamma,~n = 1,2, \ldots,N\,,\\
\nonumber
&&{\hat \Delta}_j \phi_n = \phi_{n+j}-2\phi_n+\phi_{n-j}~.
\end{eqnarray}
%--------------------------------------------------------------------
The elements $D_j(\nu)$ of the coupling term in the above equation
are as follows:
%--------------------------------------------------------------------
\begin{equation}\label{13}
D_1(\nu)=1+2\nu+3\nu^2,\; D_2(\nu)=-(\nu+2\nu^2),\; D_3(\nu)=\nu^2~. 
\end{equation}
%--------------------------------------------------------------------

We will focus on the annular JJAs, therefore, the periodic boundary
conditions will be used: $\phi_{n+N}= 2Q\pi+\phi_n$, where $Q$ is
the topological charge of the trapped soliton (fluxon).

%####################################################################
%####################################################################
\section{Linear dispersion law}
\label{disp}
%####################################################################
%####################################################################

\begin{figure}[htb]
%\sidecaption[b]
\begin{center}
\includegraphics[width=0.45\textwidth]{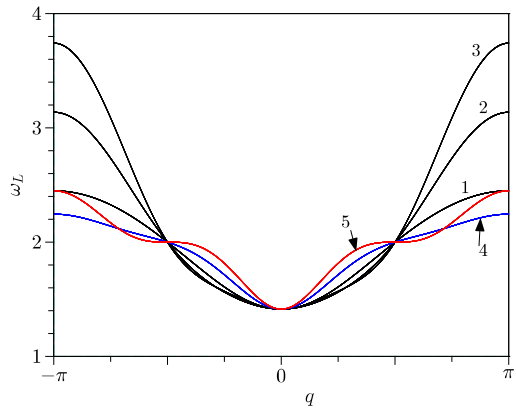}
    \caption{   (Color online). The dispersion law (\ref{14}) for the
    plane waves for $\nu=0$ (curve 1), $\nu=0.3$ (curve 2) 
    $\nu=0.5$ (curve 3), $\nu=-0.3$ (blue curve, 4) and
    $\nu=-0.5$ (red curve, 5). The rest of the parameters are
    $\eta=0.5$ and $\kappa=1$.}
   \label{fig1}   
\end{center}
 \end{figure}

The dispersion law for the small-amplitude waves (Josephson plasmons)
of Eq. (\ref{11}) can be easily obtained:
%--------------------------------------------------------------------
\begin{equation}\label{14}
\omega_L(q)=\sqrt{1+2\eta+4\kappa \left[\sum_{j=1}^3 D_j(\nu)\sin^2{
\left( \frac{jq}{2}\right) }\right ]}~. 
\end{equation}
%--------------------------------------------------------------------
The law is shown in Fig.\ref{fig1}. 
The bandwidth 
$\Delta\omega_L=\omega_L(\pi)-\omega_L(0)=
\sqrt{1+2\eta+\kappa [D_1(\nu)+D_3(\nu)]}-1$ increases as 
$\kappa$ increases. It also increases with $\nu$ if $\nu>0$.
For negative $\nu$ the opposite situation is observed and $\Delta \omega_L$
decreases, however, not monotonically because the $\nu^2$ and $\nu$ 
term can contribute differently to the final expression 
(see curves curves $4$ and $5$). 
%@@@@@@@@@@@@@@@@@@@@@@@@@@@@@@@@@@@@@@@@@@@@@@@@@@@@@@@@@@@@@@@@@@@@@
%
% Figure 1
%_____________________________________________________________________
%@@@@@@@@@@@@@@@@@@@@@@@@@@@@@@@@@@@@@@@@@@@@@@@@@@@@@@@@@@@@@@@@@@@@@
The situation of the large $|\nu|$, however, does not correspond to 
the JJA array, because 
the expansion over the powers of $\nu$ obviously fails.
Nevertheless, it may be relevant to other physical systems where them
DDbSG equation is used. It will be discussed in the next sections.

%####################################################################
%####################################################################
\section{Embedded soliton properties in the hamiltonian limit}
\label{ham}
%####################################################################
%####################################################################

In this section we consider an idealized but still very important
limit of the DDbSG equation when the dissipation and the external bias
are neglected ($\alpha=0$, $\gamma=0$). In this case we are interested 
in the existence
of the traveling wave solutions that propagate with exactly
the same shape and velocity:
%-----------------------12--------------------------------------------
\begin{equation}
 \label{12}
\phi_n(t)=\phi(n-vt)\equiv \phi(z)\,\,,\,\,z\equiv n-vt\,.
\end{equation}
%----------------------12----------------------------------------------

After substituting this ansatz into the equations of motion
(\ref{11}) one arrives to the differential equation with 
delay and advance terms:
%-----------------------x----------------------------------------------
\begin{eqnarray}\label{x}
&&v^2 \phi''(z)+\sin[\phi(z)]+\eta \sin [2\phi(z)]-\\
\nonumber
&&-\kappa \left\{ \sum_{j=1}^3 D_j(\nu)  
[\phi(z+j)+\phi(z-j)-2\phi(z)] \right \}=0~,
\end{eqnarray}
%----------------------------------------------------------------------
which can be solved only numerically. The appropriate 
pseudo-spectral method has been
developed in \cite{hmb89pd,ef90pla,defw93pd} and can trace
the traveling wave solution with a desired precision. 
Using this method we scan all possible soliton velocities
from $v=0$ to $v_{max}$. The continuous DbSG equation 
in the dimensionless variables reads 
$\phi_{tt}-v_{0}^2\phi_{xx} + V'(\phi)=0$, 
and it is Lorentz-invariant, thus $v_{max}=v_0$. Its 
discrete counterpart is not Lorentz-invariant but from the
numerical simulations we have observed that a certain maximum
soliton velocity exists. If the continuum approximation of
Eq. (\ref{11}) is performed (see the next section), one can 
find that the maximal
kink velocity is close to its continuum counterpart 
$v_{max}\sim \sqrt{\kappa\sum_{j=1}^3 j^2D_j(\nu)}$.
After scanning the whole velocity interval $[0,v_{max}]$ with the pseudo-spectral
method we observe the situation typical for the DNKG models \cite{s79prb,sze00pd,fzk99pre,kzcz02pre,acr03pd}. 
For all soliton velocities except some
discrete set $\{{\bar v}_n\}_{n=1}^M$ one obtains a bound soliton-plane
wave state with the non-vanishing oscillating tails. These solutions
are often referred to as {\it nanopterons} \cite{b90no}. 
The solutions that belong to the above-mentioned discrete set
($v={\bar v}_n$) are exponentially localized with the following
asymptotics:
%------------------------------------------------------------
\begin{equation} \label{15}
\phi(z) \to \left \{  
 \begin{array}{cc}
  0 \,,\,\, &z \to -\infty\\
  2 \pi Q \,,\,\, &z \to +\infty\,,
 \end{array}
 \right .
\end{equation}
%--------------------------------------------------------------------
where $Q$ is the topological charge. The set of these velocities
will be called {\it sliding velocities}. 
As was pointed out first in Ref. \cite{pk84pd}, 
nanopterons or solitons with non-vanishing tails appear
because for any soliton velocity $v$ there always exists a plane
wave with the same phase velocity. In other words, the
resonance condition 
%------------------------res------------------------------------------
\begin{equation}\label{res}
\omega_L(q)=vq,
\end{equation}
always has at least one real root for any $v\neq 0$. 
Appearance of the ESs means that this resonance can be avoided for the
selected set of velocities  $\{{\bar v}_n\}_{n=1}^M$.

\begin{figure}[htb]
\begin{center}
\includegraphics[width=0.45\textwidth]{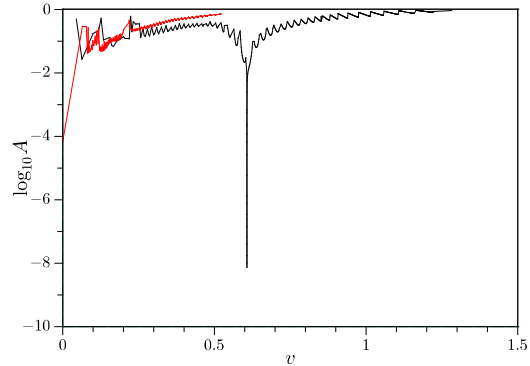}
    \caption{   (Color online). Oscillation amplitude $A$
    in the soliton tail as a function of its velocity for $\eta=0.5$,
    $\nu=-0.3$,     $\kappa=1$ (black) and     $\kappa=-0.45$ (red).  
}      
\label{fig2}
\end{center}
 \end{figure}

The typical dependence of the nanopteron tail amplitude
 $A$ on its velocity $v$ is shown in Fig. \ref{fig2}.  Existence of 
 the ES with ${\bar v}\approx 0.607$ at $\kappa=1$ and $\eta=0.5$ 
 is clearly visible.
%@@@@@@@@@@@@@@@@@@@@@@@@@@@@@@@@@@@@@@@@@@@@@@@@@@@@@@@@@@@@@@@@@@@@@
%
% Figure 2
%_____________________________________________________________________
%@@@@@@@@@@@@@@@@@@@@@@@@@@@@@@@@@@@@@@@@@@@@@@@@@@@@@@@@@@@@@@@@@@@@@ 
As it was shown for the nearest-neighbor DDbSG equation 
\cite{zs15pre},   
the existence diagram for one member of the  ES set 
has the following structure. On the parameter plane
discreteness-asymmetry ($\kappa,\eta$) there exists
a monotonous decaying 
function $\eta_c(\kappa)$, such that
%---------------------------------------------------------------------
\begin{equation}
\eta_c(\kappa)\to \left\{
\begin{array}{cc}
  0 \,,\,\, &\kappa \to 0\\
  \infty \,,\,\, &\kappa \to +\infty\,,
 \end{array}
\right .
\end{equation}
%---------------------------------------------------------------------
Everywhere below this function there is no ESs and
above this function there is at least one such soliton.
This is quite natural because the coupling $\kappa$ should be strong
enough to support the soliton propagation. Also, the parameter $\eta$
should be big enough in order to keep the system far enough
from the DSG limit that is known to have no ESs. 
This is shown in Fig. \ref{fig2} where $\kappa=0.45$
is too small to sustain an ES, but there exist an ES for $\kappa=1$.

\begin{figure}[htb]
%\sidecaption[b]
\begin{center}
\includegraphics[width=0.45\textwidth]{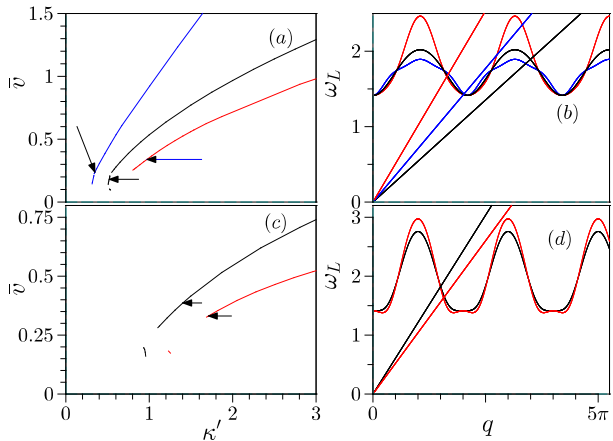}
    \caption{   (Color online).
    Dependence of the ES velocity on the renormalized 
    the coupling parameter $\kappa^\prime$ (a,c) 
    and the roots of Eq. (\ref{res}) on the dispersion laws (b,d).
    For all figures $\eta=0.5$.   
    In Fig. (a) $\nu=0$ (black), $\nu=0.3$ (red), $\nu=-0.3$ (blue). 
    In Fig. (b) the dispersion laws are shown for $\kappa=0.52$ 
    and $\nu=0$ (black), $\nu=0.3$ (red) and $\nu=-0.3$ (blue). The
    solid straight lines ${\bar v}q$ correspond to the ES velocities
    pointed by the arrows in (a) and the colors correspond to the
    different values of $\nu$ in the same way as in (a).
    Figures (c) and (d) correspond to the case when $D_3(\nu)=0$.
    In Fig. (c) $\nu=0.3$ (black) and $\nu=0.4$ (red). In Fig.  
    (d)  $\kappa=0.75$ and the color of the dispersion law and the
    ${\bar v}q$ correspond to $\nu=0.3$ (black) and $\nu=0.4$ (red).
    The straight lines correspond to the ES velocities, pointed by the arrows
    in (c). 
}
\label{fig3}   
\end{center}
 \end{figure}

We focus on the dependence of the ES velocity (sliding velocity) ${\bar v}$
on the coupling constant $\kappa$ for the fixed value $\eta$.
For this purpose we construct the dependence of  ${\bar v}$
  on the renormalized coupling constant
%--------------------------------------------------------------------
\begin{equation}
\kappa'=D_1(\nu) \kappa = (1+2\nu+3\nu^2) \kappa~. 
\end{equation}
%--------------------------------------------------------------------
This constant should be used instead of $\kappa$ because
it is the correct measure of the nearest-neighbor interaction
if $\nu\neq 0$ [see Eq. (\ref{11})]. Thus, by comparing the ${\bar v}(\kappa')$ for
the different values of $\nu$ we find out how the next-to-nearest
and second-to-nearest 
neighbor interaction influences the existence of embedded
solitons. In Fig. \ref{fig3}(a) the dependence of the ES velocity
on $\kappa'$ is given. In the case of $\nu<0$ the next-neighbor 
interactions increase the existence range of ESs and their velocity
while for $\nu>0$ the existence range together with the velocity
decrease. The ${\bar v}(\kappa')$ dependence is not always purely
monotonic and may consist of several pieces, as for $\nu=0$ and 
$\nu=-0.3$ 
%@@@@@@@@@@@@@@@@@@@@@@@@@@@@@@@@@@@@@@@@@@@@@@@@@@@@@@@@@@@@@@@@@@@@@
%
% Figure 3
%_____________________________________________________________________
%@@@@@@@@@@@@@@@@@@@@@@@@@@@@@@@@@@@@@@@@@@@@@@@@@@@@@@@@@@@@@@@@@@@@@

This can be easily
understood from the analysis of the linear spectrum of the array.
It is important to recall the result of Ref. \cite{acr03pd} where the
lowest bond of the ES velocity has been determined. This paper states
that ${\bar v}$ cannot lie in the parameter range where the 
Eq. (\ref{res}) has more than one root. As the velocity is decreased,
the $q{\bar v}$ line can cross the linear spectrum band $\omega_L$
three, five or more times. This argument originates from the idea that
the ES appears as a result of the destructive interference of the plane waves,
emitted by the moving solitons. It was first formulated in Ref. 
\cite{pk84pd} for the bound state of two or more $0-2\pi$ kinks
in the DSG equation. In the DDbSG equation there are two limits when
it turns into the standard DSG equation. The limit $\eta\to 0$ is trivial.
Another limit is $\eta \to \infty$. In this case the proper renormalization 
of the nonlinear term is necessary (see Ref. \cite{nknfh10pc}). In this
limit the term $\sin 2\phi$ will dominate over the $\sin \phi$ term,
and, as a result, the $0-\pi$ kinks will replace the $0-2\pi$ kinks. 
Thus, as was also pointed out in Ref. \cite{bkm01wm} for the 
strongly dispersive continuous DbSG, 
in the intermediate 
situation with $\eta$ being large enough but finite, one can speak about the $0-2\pi$
kinks as weakly coupled pair of two $0-\pi$ kinks. Therefore, the
ES for $0<\eta<\infty$ is a continuation of the two $0-\pi$ kink bound state 
from the limit $\eta=\infty$. Of course, if $\eta$ decreases down to
the critical value $\eta_c$ this bound state breaks down 
and we have no ES.
The destructive interference between these two ``virtual kinks''
will work only if there is just one resonance (\ref{res}) to be avoided. Therefore,
if the $vq$ line crosses the dispersion law more times, additional resonances
appear and they cannot be suppressed. 

In Fig. \ref{fig3}(b) this argument can be clearly demonstrated as we
show how the resonance condition (\ref{res}) works for the ES 
near the edge of its existence
area. The dispersion laws corresponds to $\kappa=0.52$ in all three cases.
The $\nu=0$ case is shown by the black line. The corresponding 
$\bar v(\kappa')$ dependence is fragmented, and, apart from the main
curve, has two small pieces below it. The ES velocity value, that 
 corresponds to the black line [${\bar v}\approx 0.18098$, pointed by the
 arrow in Fig. \ref{fig3}(a)] creates one
 crossing with the dispersion law in the fourth Brillouin zone (BZ). 
For larger $\kappa'$ and larger velocity the respective crossing would occur
in the second BZ and that would correspond to the main curve of the $\bar v(\kappa')$
dependence. Precisely this is shown by two other resonances $\nu=-0.3$
(blue curve) and $\nu=0.3$ (red curve). In these two cases the dispersion
law is crossed by the respective ${\bar v}q$ line in the second BZ. For the 
$\nu=-0.3$ the the slope of the ${\bar v}q$ line (blue) crosses the
dispersion law once but is very close to the
situation when it will cross the dispersion law three times. Thus, there
exists a small forbidden interval for velocities where no ES is possible.
After passing this interval there is again one root of Eq. (\ref{res}), but
in the fourth BZ. Depending on the width of the spectrum, $\Delta \omega_L$,
there could be single roots of Eq. (\ref{res}) in the next even BZs.
Therefore, we observe that positive $\nu$ plays a destructive role
in the ES formation because it causes widening of the linear wave band, and,
as a result, reduces the parameter space for the one-root solutions of
the resonance equation (\ref{res}). For the same reason, the case 
of $\nu<0$ is more 
favorable for the existence of ESs, as the linear spectrum becomes more narrow
in comparison to the $\nu=0$ situation. 
As an interesting side observation, we note that the ES velocity always corresponds to the
root of (\ref{res}) that lies in the even BZ. From the plots it is quite
obvious that a single root is not possible in the $3,5,\ldots,(2n+1)$th BZ.
However we have not seen any ES that corresponds to the resonance in the first
BZ either. We think that ES can exist only if the respective group
velocity is negative because then the radiated energy travels backwards
and the soliton can separate itself from it.

To study further the influence of the linear wave spectrum on the ES
existence we consider the case when the dispersion law
has a local maximum at $q=0$ and its minimum is placed between
$q=0$ and $q=\pi$, and, in addition to that 
$\omega_L(q_{min})<\sqrt{1+2\eta}$. 
This can be achieved if the
 term $D_2$ in (\ref{13}) dominates over the first term $D_1$.
It should be noted that this situation can not be achieved if
the expansion (\ref{8}) holds. Thus, it is not directly applied to the
JJ array. 
We put by hands $D_3=0$ and take $\nu=0.3$ and $\nu=0.4$. In this case the
dispersion law takes the shape as in Fig. \ref{fig3}(d). 
In Fig. \ref{fig3}(c) we observe the further decrease of the existence
area of the ES on the $\kappa$ axis. For $\nu=0.5$ we did not manage
to find ES at all and for $\nu=0.6$ did not find any traveling-wave
solutions of Eq. (\ref{x}). This particular case is
different from the dispersion laws discussed in the previous
paragraphs because its lower bond decreases as $\kappa$ increases.
At some point $\omega_L(q_{min})$ will reach zero, thus, signaling 
instability. This is not surprising if we think for a moment about 
Eq. (\ref{11}) as a lattice of interacting particles and $\phi_n$'s
as their spatial coordinates. Then the nearest-neighbor interaction 
term describes attractive interaction because $D_1(\nu)>0$ and 
the next-to-nearest interaction is repulsive because $D_2(\nu)<0$. 
If $\nu$ is sufficiently large, the attractive interaction is no 
longer strong enough
to balance the repulsive term and the whole system becomes 
unstable. Even in the parameter range where it stays stable, 
the linear spectrum width becomes so big that it becomes impossible
to have one root of the resonance equation (\ref{res}) anywhere 
except the first BZ.

%####################################################################
%####################################################################
\section{Current-voltage characteristics of the annular array}
\label{cvc}
%####################################################################
%####################################################################

Realistic simulations of the JJAs should take into account the
effects of dissipation that originate from the normal electron
tunneling across each junction. Also, the external DC bias should be
included into consideration. Thus, the full 
Eqs. (\ref{11}) should be solved with $\alpha>0$ and $\gamma\neq 0$.

\subsection{Continuous approximation}

In the continuous approximation the DDbSG equation can be written
as a standard double sine-Gordon (DbSG) equation:
%----------------------------------------------------------------------
\begin{eqnarray}\label{16}
&&\phi_{tt}-v_0^2 \phi_{xx}+\sin \phi + \eta \sin{2\phi}=-\alpha \phi_t+
\gamma~,\\
\nonumber
&&v_0^2=\kappa \sum_{j=1}^3 j^2 D_j(\nu)~.
\end{eqnarray}
%----------------------------------------------------------------------
In order to get the current-voltage characteristics (CVCs) in this case 
we  use the energy balance approach, developed previously \cite{ms78pra}. 
According to this approach the total power $\bar {V} \gamma$, 
applied to the soliton is compensated by the dissipation,
$\bar {V}_c \gamma=-P_{diss}$. The dissipative losses can
be easily computed if we use the exact solution \cite{cps86pd,cgm83prb}
of the 
unperturbed continuous DbSG equation with the arbitrary velocity $v$:
%----------------------------------------------------------------------
\begin{eqnarray}\label{20}
 &&P_{diss}=-\alpha \int_{-\infty}^{+\infty}\phi_t^2 dx=
 \frac{4\alpha \Phi(\eta)}{\pi \sqrt{v_0^2-v^2}},\\
 \label{21}
 &&\Phi(\eta)= \frac{\sqrt{1+2\eta}}{2}\left [1+ \right.
\nonumber  \\
 &&\left. +\frac{1}{\sqrt{2\eta(2\eta+1)}}~
\mbox{arctanh} \sqrt{\frac{2\eta}{1+2\eta}}\right ]~,
\end{eqnarray}
%----------------------------------------------------------------------
%The voltage drop is connected with the soliton velocity as 
%$\bar{V}_c=\frac{2 \pi v_\infty}{N}$ 
As a result we arrive to the 
analytical expression for the average voltage drop:
%----------------------------------------------------------------------
\begin{equation}\label{22}
\bar{V}_c=\frac{2 \pi v}{N}
=\frac{2 \pi v_0}{N}\left [1+ \Phi^2(\eta)
\left ({4\alpha} \over {\pi\gamma}\right)^2 \right ]^{-1/2}~.
\end{equation}
%----------------------------------------------------------------------
In the limit $\eta\to 0$ the standard SG equation is restored.
From Eq. (\ref{21}) one observes that $\Phi(\eta)_{\eta\to 0}\to 1$.
Thus, the equilibrium velocity coincides with the 
equilibrium velocity of the SG equation 
$v=\left [1+\left 
({4\alpha} \over {\pi\gamma}\right)^2\right ]^{-1/2}$
\cite{ms78pra}.

%#####################################################################
\subsection{Numerically computed current-voltage characteristics}
%#####################################################################

\begin{figure}[htb]
%\sidecaption[b]
\begin{center}
\includegraphics[width=0.5\textwidth]{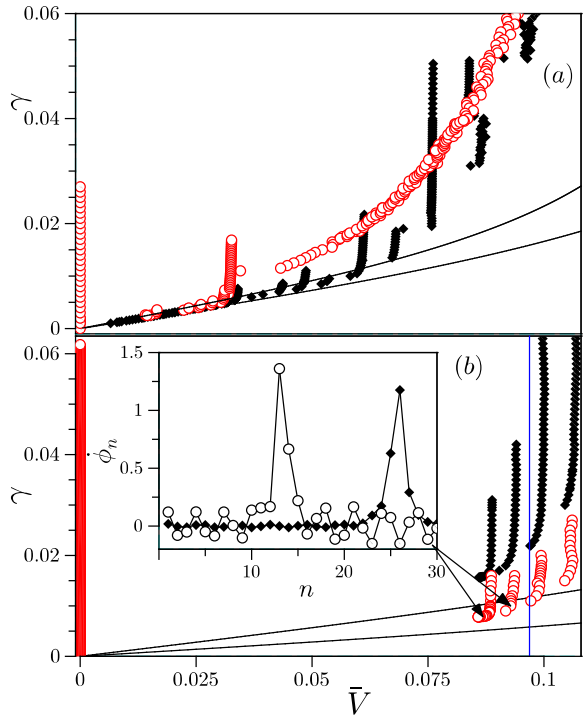}
    \caption{   (Color online).
    Current-voltage characteristics for the annular JJA with $N=30$
    junctions. Figure (a) corresponds to  
    $\kappa=0.5$, $\nu=\eta=0$ (red $\circ$) and $\eta=0.1$, $\kappa=0.8$
    (black $\blacklozenge$).
    In both cases  $\alpha=0.02$ and $\nu=0$. 
    Continuous approximation is shown by the solid black lines. 
    Figure (b) corresponds to $\kappa=0.8$, $\eta=0.5$, $\nu=-0.3$, 
    $\alpha=0.02$ (black $\blacklozenge$) and $\alpha=0.01$ (red $\circ$).
   Inset shows the Josephson phase distribution
    for $\gamma=0.01$ ($\circ$) and $\gamma=0.0079$ ($\blacklozenge$). 
    Vertical blue line corresponds to the voltage $V=2\pi {\bar v}/N$, 
    where ${\bar v}$ is the ES velocity (see text for details).
}
\label{fig4}   
\end{center}
 \end{figure}

The CVCs provide the necessary information
about the JJ array dynamics and are accessible through experimental 
measurements. The average voltage drop is defined as
%----------------------------------------------------------------------
\begin{equation}\label{24}
 {\bar V}= \frac{1}{N}\sum_{n=1}^N \lim_{t
\rightarrow \infty} \frac{1}{t}\int_0^t {\dot \phi}_n(t') dt'~.
\end{equation}
%----------------------------------------------------------------------
If there is a soliton that moves along the array with velocity $v$ it
will produce the average voltage drop $2\pi v/N$.
Since Eq.~(\ref{11}) is dissipative, we are going to deal
with its attractor solutions.
The numerically computed CVC curves are shown in Fig. \ref{fig4}. 
To obtain these 
figures we have changed the bias current $\gamma$ in both directions:
from $\gamma=0$ till $\gamma=0.06$ and in the reverse way. 
To integrate Eqs. (\ref{11}) the 4th order Runge-Kutta method was used.

We remind here the basic difference between the CVCs in the continuous
JJs and in the JJ arrays. In the former case the CVC is a 
continuous monotonic
function given by Eq. (\ref{22}). The soliton dynamics in the  
continuous JJ is qualitatively similar to the particle moving in the
viscous liquid under the influence of gravitation, where the DC
bias plays the role of gravitation and $\alpha$ plays the 
role of viscosity. Discreteness and periodic boundary conditions
bring the fundamental changes to the shape of the CVCs. In the JJ array
they constitute a series of separate curves, as can be seen in 
Fig. \ref{fig4}. The presence of the periodic boundary conditions
means that only a certain integer number of the plane wave
periods can fit into the array of $N$ junctions \cite{ucm93prb,bhz00pre}.
Each branch corresponds to the distinct number of periods and
the wavelength of each period is defined by the resonance 
condition (\ref{res}). Sometimes it is not possible to identify 
separate branches, especially for the large values of $\gamma$, because
the JJA dynamics may be quasiperiodic or even chaotic.
The continuous CVCs (\ref{22}) are given by the solid lines. It appears that this
approximation is in satisfactory agreement with the numerical
data only near the origin of the CVC.
Another prominent signature of discreteness is hysteresis of the CVCs. 
If we start from the superconducting state (pinned soliton), it will stay
pinned until the bias $\gamma$ reaches the critical value. This critical value
depends only on the static properties of the array and does not
depend on dissipation. The retrapping current is the minimal current
for which soliton motion is possible. This current decreases with
the decrease of $\alpha$.

The case when no ESs exist in the hamiltonian limit is presented in
Fig. \ref{fig4}(a). One can observe that the CVCs occupy almost all
accessible voltage range. For $\kappa=0.5$ the CVC is approximately
continuous for the larger voltages, while for the smaller voltages
some vertical branches can be identified. Closer to the origin the 
separate branches can hardly be distinguished from each other. 
For $\kappa=0.8$ the vertical branches appear more clearly because for
the larger values of $\kappa$ the JJ dynamics is less chaotic. The
behavior near the origin is similar to the $\kappa=0.5$ case.
%@@@@@@@@@@@@@@@@@@@@@@@@@@@@@@@@@@@@@@@@@@@@@@@@@@@@@@@@@@@@@@@@@@@@@
%
% Figure 4
%_____________________________________________________________________
%@@@@@@@@@@@@@@@@@@@@@@@@@@@@@@@@@@@@@@@@@@@@@@@@@@@@@@@@@@@@@@@@@@@@@
The best manifestation of the ESs is possible if the parameter $\eta$ 
is large enough to keep the system far from the DSG limit. 
In Fig. \ref{fig4}(b) the case of $\eta=0.5$ is considered. On the
respective CVC one observes the situation similar to the previously
discussed  case of the DDbSG model with just the
nearest neighbour interactions ($\nu=0$) \cite{zs15pre}. 
The general picture looks more ordered with the distinct separate 
almost vertical CVC branches that exist only above some 
certain value of ${\bar V}$. 
The {\it inaccessible voltage interval} (IVI) $[0,V_{IVI}]$ is formed,
within which there is no CVC branches. The upper edge of this range
appears to lie close to the voltage, produced by the ES in the
hamiltonian limit, which is ${\bar V}=2\pi {\bar v}/N \approx 0.0969$ and
is shown in the figure by the vertical blue line. 
For the respective 
system parameters ($\eta=0.5$,
$\nu=-0.3$) the upper bond of the IVI is $V_{IVI}\approx 0.086$, and
it weakly depends on dissipation (up to the fourth decimal for
$\alpha=0.01$ and $\alpha=0.02$). From Fig. \ref{fig4} we see that 
not the lowest but
the third lowest branch of CVC springs off the voltage 
$2\pi {\bar v}/N$ that corresponds to the ES in the hamiltonian limit.
The threshold value of the DC bias at the voltage $V_{IVI}$ decreases
as $\alpha$ tends to zero, what is a principal difference from the
situation when no ESs is possible. The soliton profile at the edge
of the IVI has very small radiating tails 
[see the inset of Fig. \ref{fig4}(b)] while the soliton profile from
the neighboring branch of the CVC has much better pronounced oscillating
asymptotics.
We have computed the detuning $\epsilon=2\pi {\bar v}/N-V_{IVI}$ of the
IVI edge from the voltage created by the moving ES. This detuning parameter
is given in Tab. \ref{tb1} for the parameters of Fig. \ref{fig4} except
$\nu$ which is increased up to $\nu=0$.
%@@@@@@@@@@@@@@@@@@@@@@@@@@@@@@@@@@@@@@@@@@@@@@@@@@@@@@@@@@@@@@@@@@@@@@
\begin{table}[htb]
\begin{tabular}{c|c}
\hline
\hline
$\nu$ & $\epsilon$      \\
\hline      
  -0.3               & 0.0109      \\
  -0.2                & 0.0063      \\
  -0.1               & 0.0045      \\
  -0.05              & 0.0047      \\
   0                 &  0.0008           \\
\hline  
\hline
\end{tabular}
\caption{The detuning parameter $\epsilon=2\pi {\bar v}/N-V_{IVI}$   
as a function of $\nu$. The other parameters are as in Fig.\ref{fig4}}
\label{tb1}
\end{table}
%@@@@@@@@@@@@@@@@@@@@@@@@@@@@@@@@@@@@@@@@@@@@@@@@@@@@@@@@@@@@@@@@@@@@@@
As we approach to the pure next-neighbor limit ($\nu$), the difference 
between the upper edge of IVI and the voltage, produced by the moving
ES becomes very small.
We conclude that in the presence of small dissipation and for the small
values of DC bias $\gamma$ the JJA dynamics settles on the 
attractor that originates from  the ES soliton in the hamiltonian limit.

%####################################################################
%####################################################################
\section{Discussion and conclusions}\label{conc}
%####################################################################
%####################################################################

In this paper we have investigated how the presence of the 
next-to-nearest and second-to-nearest interactions in the nonlinear
discrete Klein-Gordon (NDKG) lattice influences the properties of the
embedded solitons (ES). We have taken the discrete double sine-Gordon
equation as a working model because of the broad range of its application
in various fields of modern physics and because it is known to
support ESs in the limit of the nearest-neighbor interactions \cite{zs15pre}. 
In particular, this equation is used for modeling of the 
arrays of JJs, in particular arrays of SFS or SIFS
junctions \cite{gkkb07prb}. The appearance of the next-to-nearest and 
second-to-nearest 
interactions is due to the fact that the inductive coupling between
the neighboring cells of the array was taken into account. 
If this coupling is weak comparing
to the self-inductance of the cell, the resulting equations of
motion for the Josephson phase can be rewritten as the DDbSG equation with
the next-neighbor and second-neighbor interactions. 
The interaction with the $j$th neighbor
comes as a discrete Laplasian term with the coefficient of the order 
${\cal O}(\nu^{j-1})$, where $\nu$ is the ratio of the mutual inductance
between the neighboring cells of the array to the self-inductance of the
cell. 

We have demonstrated that existence of embedded solitons (ESs) depends 
primarily on the properties of the spectrum of the linear waves
of the system. Our results confirm previous findings \cite{acr03pd}
where it was shown that a discrete ES cannot exist with velocities
for which Eq. (\ref{res}) has more than one root. The case of $\nu<0$
is more relevant to the JJ physics and for it  
the next-to-nearest and the second-to-nearest interactions create
more favorable conditions for the ESs formation as compared to the
$\nu=0$ limit. In particular,
the existence range (in the terms of the nearest-neighbor
interaction term) for the ESs on the increases, the
ES velocity increases as well. For the positive values of $\nu$ the
situation is opposite. The ES velocity decreases, as well as the existence
range. The explanation is as follows: in the former case the linear wave
spectrum narrows, thus creating more possibilities for having just one
root of the resonance condition (\ref{res}). In the latter case the
linear band widens and, as a result, makes more difficult to have just
one root of this equation.

This research can be further extended into other physical models that
are not connected to the Josephson effect. Recent research on the 
nonlinear electric circuits with the next-neighbor interactions 
\cite{cake18pre} seems to be a
promising field for application of the ideas developed in this
article.

%####################################################################
%####################################################################
\section*{Acknowledgements}
%####################################################################
%####################################################################

Publication is based on the research provided by the 
 grant support of the State Fund For
Fundamental Research (project No. F76/6-2018).

%\bibliographystyle{apsrev}
%\bibliography{/home/yzolo/TEX/BIB/books,/home/yzolo/TEX/BIB/jj,/home/yzolo/TEX/BIB/db,/home/yzolo/TEX/BIB/acou,/home/yzolo/TEX/BIB/kink}

\end{document}